\documentclass[12pt]{article}

\usepackage[utf8]{inputenc}
\usepackage{geometry}
\usepackage{graphicx}
\usepackage{array}
\usepackage{subfig}
\usepackage{slashed}
\usepackage{amsmath}
\usepackage{amsfonts}
\usepackage{amssymb}
\usepackage[cm]{fullpage}
\usepackage{cite}
\usepackage{epsfig}
\usepackage{comment}
\usepackage{hyperref}
\usepackage{tikz}
\usepackage{feynmp}
\usepackage{cancel}
\usepackage{mathrsfs}
\usepackage{mathtools}

\usepackage{cleveref}
\crefname{section}{§}{§§}
\Crefname{section}{§}{§§}

\numberwithin{equation}{section}

\def\ip{${\mathscr I}^+$}

\def\e{{\epsilon}}
\def\cs{${\cal S}$}

 \def\p{\partial}
 \def\bz{{\bar z}}
 \def\bw{{\bar w}}
 
\def\0{{(0)}}
\def\1{{(1)}}
\def\2{{(2)}}

\def\ci{{\mathscr I}}

\def\<{\langle }
\def\>{\rangle }
\def\bw{{\bar w}}
\newcommand{\bea}{\begin{eqnarray}}
\newcommand{\eea}{\end{eqnarray}}
\newcommand{\be}{\begin{equation}}
\newcommand{\ee}{\end{equation}}
\newcommand{\ba}{\begin{align}}
\newcommand{\ea}{\end{align}}

\renewcommand{\epsilon}{\varepsilon}

   \makeatletter
  \let\over=\@@over \let\overwithdelims=\@@overwithdelims
  \let\atop=\@@atop \let\atopwithdelims=\@@atopwithdelims
  \let\above=\@@above \let\abovewithdelims=\@@abovewithdelims
\renewcommand\section{\@startsection {section}{1}{\z@}%
                                   {-3.5ex \@plus -1ex \@minus -.2ex}%nn
                                   {2.3ex \@plus.2ex}%
                                   {\normalfont\large\bfseries}}

\renewcommand\subsection{\@startsection{subsection}{2}{\z@}%
                                     {-3.25ex\@plus -1ex \@minus -.2ex}%
                                     {1.5ex \@plus .2ex}%
                                     {\normalfont\bfseries}}

\newcommand{\bra}[1]{\left< #1 \right|}
\newcommand{\ket}[1]{\left| #1 \right>}

\newcommand{\beq}{\begin{equation}}
\newcommand{\eeq}{\end{equation}}
\newcommand{\beqa}{\begin{eqnarray}}
\newcommand{\eeqa}{\end{eqnarray}}
\newcommand{\beqar}{\begin{eqnarray*}}

\newcommand{\Or}{\mathcal{O}}

\newcommand{\cm}{{\cal M}}
\newcommand{\ve}{{\varepsilon}}

\newcommand{\F}{{\cal F}}
\newcommand{\A}{{\cal A}}
\def\[{\big[}
\def\]{\big]}

\def\del{\partial}

\newcommand{\bd}[1]{\begin{fmffile}{#1}\begin{fmfgraph*}}
\newcommand{\ed}{\end{fmfgraph*}\end{fmffile}}

\begin{document}
\begin{titlepage}
\unitlength = 1mm
\ \\
\vskip 3cm
\begin{center}

{\LARGE{\textsc{New Symmetries of  Massless QED}}}

\vspace{0.8cm}
Temple He, Prahar Mitra, Achilleas P. Porfyriadis and Andrew Strominger

\vspace{1cm}

{\it  Center for the Fundamental Laws of Nature, Harvard University,\\
Cambridge, MA 02138, USA}

\vspace{0.8cm}

\begin{abstract}

An infinite number of physically nontrivial symmetries are found for abelian gauge theories with massless charged particles. They are generated by large $U(1)$ gauge transformations that asymptotically approach an arbitrary function $\e(z,\bz)$ on the conformal sphere at future null infinity (\ip) but are independent of the retarded time.  The value of $\e$ at  past null infinity ($\ci^-$) is determined from that on \ip\ by the condition that it take the same value at either end of any light ray crossing Minkowski space. The $\e\neq$ constant symmetries are spontaneously broken in the usual vacuum. The associated Goldstone modes are zero-momentum photons and comprise a $U(1)$ boson living on the conformal sphere. The Ward identity associated with this asymptotic symmetry is shown to be the abelian soft photon theorem.

 \end{abstract}

\vspace{1.0cm}

\end{center}

\end{titlepage}

\pagestyle{empty}
\pagestyle{plain}

\def\vx{{\vec x}}
\def\p{\partial}
\def\po{$\cal P_O$}

\pagenumbering{arabic}

\tableofcontents

\section{Introduction}\label{sec:intro}

Recently a general equivalence relation has emerged between soft theorems and asymptotic symmetries \cite{as,as1,hms,Cachazo:2014fwa,Casali:2014xpa, Schwab:2014xua,Bern:2014oka,He:2014bga,larjoski,Cachazo:2014dia,jeddi,Adamo:2014yya,Geyer:2014lca,Kapec:2014opa,schwab,Bianchi:2014gla,Balachandran:2014hra, Broedel:2014fsa,Bern:2014vva}. Soft theorems are relations between $n$ and $n+1$ particle scattering amplitudes, where the extra particle is soft.  Any linear relation between scattering amplitudes can be recast as an infinitesimal symmetry of the \cs-matrix.  It is gratifying that in some cases  the resulting symmetries have turned out to be known space-time or gauge symmetries. For example Weinberg's soft graviton theorem  \cite{steve,stevebook} is equivalent to a symmetry of the \cs-matrix  generated by a certain diagonal subgroup  \cite{as1} of the product of BMS \cite{bms} supertranslations acting on past and future null infinity, \ip\ and $\ci^-$.

This  equivalence relation is of interest for several reasons. It ``explains'' why soft theorems exist and are so universal: they arise from a symmetry principle. Moreover, it imparts observational meaning to Minkowskian asymptotic symmetries, which have at times eluded physical interpretation. The framework has proven useful for establishing new symmetries \cite{Kapec:2014opa} and  new soft theorems \cite{Cachazo:2014fwa,Casali:2014xpa, Schwab:2014xua}. In the quantum gravity case, the symmetries provide the starting point for any attempt at a holographic formulation, see $e.g.$ \cite{ash}. In the gauge theory case, they are potentially useful for improving  the accuracy of collider predictions, see $e.g.$ \cite{white}.

The purpose of the present paper is to argue that the soft photon theorem in massless QED \cite{Yennie:1961ad}\cite{steve}\cite{Grammer:1973db} can be understood as a new asymptotic symmetry. The symmetry is generated by ``large'' $U(1)$ gauge transformations which approach  an arbitrary function $\e(z,\bz)$ on the conformal sphere at $\ci$ but are constant along the null generators, even as they antipodally cross from $\ci^-$ to \ip\ through spatial infinity. Except for  the constant transformation, these symmetries are spontaneously broken in the conventional vacuum. The soft photons appear as Goldstone modes living on the sphere at the boundary of $\ci$.

These large $U(1)$ gauge symmetries are precise analogs of BMS supertranslations in gravity.\footnote{It would be interesting to systematically derive this large $U(1)$ symmetry group using the type of asymptotic analysis employed in BMS \cite{bms}: herein the non-triviality of the symmetries is established by their equivalence to the soft photon theorem.}  It is curious  that discovery of the gravitational symmetry preceded its electromagnetic analog by a half-century.

The relation between soft theorems and asymptotic symmetries of \ip\ (but not of the \cs-matrix), was described already in \cite{as}, which in turn was inspired by \cite{juan}.  Two ``simplifying'' restrictions were made in the analysis of \cite{as}: the incoming state was required to be invariant under the large gauge symmetries, and the parameter $\e(z,\bz)$ was required to be locally holomorphic. However, far from simplifying the analysis, these restrictions obscured the underlying structure. The present analysis both simplifies and generalizes that of \cite{as}.

This paper considers theories in which there are no stable massive charged particles, and the quantum state begins and ends in the vacuum at past and future timelike infinity.  Of course, in real-world QED the electron is a stable massive charged particle, so it is highly desirable  to generalize our analysis to this case.\footnote{The present analysis is relevant to hard scattering in QED when the electron mass becomes negligible.} However, stable massive charges create technical complications because the charge current has no
flux through future null infinity.  Rather, there is charge flux across timelike infinity which becomes a singular point in the conformal compactification of Minkowski space. In principle a systematic treatment of this singularity should be possible -- the fields disperse and are weakly interacting; nonetheless, this is well beyond the scope of the present paper.

This paper is organized as follows. In section \ref{sec:asexp1} we describe the classical final data formulation at \ip. Section \ref{sec:largegauge} gives the asymptotic symmetries which are the QED analogs of BMS symmetries and constructs the associated charges. In section \ref{sec:canform} the commutators at \ip\ are given and the charges of section 3 are shown to generate the symmetries. This requires a careful treatment of the Goldstone modes and boundary conditions at the boundaries of \ip. Section \ref{secim} gives the corresponding formulae for $\ci^-$. In section \ref{sec:matching} we give conditions which tie the data of $\ci^-$ to that of \ip\ and thereby defines the scattering problem. The conditions are shown to break the separate asymptotic symmetries to a diagonal subgroup preserving the \cs-matrix.  In section \ref{sec:quantum} the quantum Ward identity of this symmetry is shown to relate scattering amplitudes with and without a soft photon insertion. Finally in section \ref{sec:softphoton} we show that this Ward identity is the soft photon theorem.

\section{Asymptotic expansion at $\ci^+$}\label{sec:asexp1}

In this subsection we consider the canonical final data formulation of $U(1)$ electrodynamics coupled to massless charged matter at future null infinity ($\ci^+$). It is convenient to adopt retarded coordinates
\begin{equation}\label{retardedcoord}
	ds^2 = -du^2 - 2du dr + 2r^2\gamma_{z\bz}dz d\bz,
\end{equation}
where $u = t - r$ and $\gamma_{z\bz} = \frac{2}{(1+z\bz)^2}$ is the round metric on the conformal sphere.\footnote{We have set $z = e^{i\phi}\tan\frac{\theta}{2}$.}  Thus $\ci^+$ is the null $S^2 \times {\mathbb R}$ boundary at $r = \infty$ with coordinates $(u,z,\bz)$. $\ci^+$ has boundaries at $u = \pm \infty$, which we denote $\ci^+_\pm$.

The bulk equations of motion for $U(1)$ gauge theory are
\begin{equation}
\begin{split}\label{eqofmotion}
\nabla^\nu \F_{\nu\mu} = e^2 j^M_\mu  ,
\end{split}
\end{equation}
where $\F_{\mu\nu} = \p_\mu \A_\nu - \p_\nu \A_\mu$ and $j_\mu^M$ is the conserved matter obeying $\nabla^\mu j^M_\mu = 0$. The equations \eqref{eqofmotion} have a gauge symmetry under which
\begin{equation}
\begin{split}
	\delta_{\hat \ve} \A_\mu = \p_\mu {\hat \ve},
\end{split}
\end{equation}
with periodicity
\begin{equation}
\begin{split}
	{\hat \ve} \sim {\hat \ve} + 2 \pi .
\end{split}
\end{equation}
We work in the retarded radial gauge
\begin{align}
\mathcal A_r &= 0 , \label{gauge_fix1} \\
	\left.\A_u\right|_{\ci^+} &= 0 \label{gauge_fix2}.
\end{align}

We wish to expand the fields around $\ci^+$. The radiation flux through $\ci^+$ is proportional to $\int_{\ci^+} \F_u{}^z \F_{uz } $. To ensure that the radiation flux is nonzero and finite, we require $\A_z \sim \Or(1)$ near $\ci^+$. Following \eqref{gauge_fix2}, we also require $\A_u \sim \Or(1/r)$ near $\ci^+$, giving the expansion
\begin{equation}\label{fieldexpand}
\begin{split}
	\A_z(r,u,z,\bz) &= A_z(u,z,\bz) + \sum_{n=1}^\infty \frac{A_z^{(n)}(u,z,\bz)}{r^n} , \\
	\A_u(r,u,z,\bz) &= \frac{1}{r}A_u(u,z,\bz) + \sum_{n=1}^\infty \frac{A_u^{(n)}(u,z,\bz)}{r^{n+1}}.
\end{split}
\end{equation}
The leading terms in the field strengths near $\ci^+$ are then $\F_{z\bz} = \Or(1)$, $\F_{ur} = \Or(r^{-2})$, $\F_{uz} = \Or(1)$, and $\F_{rz} = \Or(r^{-2})$ with coefficients
\begin{equation}
\begin{split}
	F_{z\bz} &= \p_zA_{\bz} - \del_{\bz}A_z , \\
	F_{uz} &= \p_uA_z  , \\
	F_{rz} &= -A_z^{(1)} , \\
	F_{ur} &= A_u.
\end{split}
\end{equation}
Note that the fields $F$ and $A$ live on $\ci^+$ and have no $r$ dependence. Substituting \eqref{fieldexpand} into the $\mu = u$ component of \eqref{eqofmotion}, we get the leading constraint equation
\begin{equation} \label{eom}
	\gamma_{z\bz} \p_u A_u = \p_u \left( \p_z A_\bz + \p_\bz A_z \right) + e^2 \gamma_{z\bz}  j_u ,
\end{equation}
where
\begin{equation}
\begin{split}
j_u (u,z,\bz) = \lim_{r\to\infty} \left[ r^2 j_u^M (u,r,z,\bz) \right].
\end{split}
\end{equation}
We will be interested in configurations which revert to the vacuum in the far future, i.e.
\begin{equation}
\begin{split}\label{futvac}
F_{ur} |_{\ci^+_+} = F_{uz} |_{\ci^+_+} = 0.
\end{split}
\end{equation}
From \eqref{eom} and \eqref{futvac} we can determine $A_u$ in terms of $A_z$ and $A_\bz$ (for a given $j_u$), which we will take to be coordinates on the asymptotic phase space $\Gamma^+$. Subleading terms in the $\frac{1}{r}$ expansions of all the other equations of motion then determine the expansion of $\A$ in terms of the final data $A_z$ and matter current.

The analogous structure at $\ci^-$ is described in section \ref{secim} below.

\section{Large gauge transformations}\label{sec:largegauge}

The gauge conditions \eqref{gauge_fix1} and \eqref{gauge_fix2} leave unfixed residual gauge transformations generated by an arbitrary function approaching ${\hat \ve} = \ve(z,\bz)$ on the conformal sphere at $r=\infty$. We will refer to these as ``large gauge transformations.''The action on $\Gamma^+$ is
\begin{equation}
\begin{split}
	\delta_\ve A_z (u,z,\bz) = \del_z\ve (z,\bz).
\end{split}
\end{equation}
These comprise the asymptotic symmetries considered in this paper. The charge that generates this transformation can be determined by Noether's procedure
\begin{align}\label{charge}
Q^+_{\ve} &= \frac{1}{e^2} \int_{\ci^+_-}d^2 z \gamma_{z\bz} \ve F_{ru} = \frac{1}{e^2} \int_{\ci^+} du d^2 z \ve \left[   \p_u \left( \p_z A_\bz + \p_\bz A_z \right) + e^2 \gamma_{z\bz} j_u \right] .
\end{align}
In the second equality we have integrated by parts, assumed the final charge relaxes to zero at $\ci^+_+$ and used the constraint \eqref{eom}.  For the special case $\ve=1$, $Q^+_1$ is the total initial electric charge which obeys
\begin{equation}
\begin{split}\label{mattercharge}
	Q^+_1 = \int_{\ci^+} du d^2 z\gamma_{z\bz} j_u.
\end{split}
\end{equation}
For the choice $\ve (z,\bz) = \delta^2(z - w)$ one has the fixed-angle charge
\begin{equation}
\begin{split}
	Q^+_{w\bw} = \frac{1}{e^2} \int_{-\infty}^\infty du \left[   \p_u \left( \p_w A_{\bw} + \p_\bw A_w \right) + e^2 \gamma_{w\bw} j_u \right] .
\end{split}
\end{equation}
This is the total outgoing electric charge radiated into the fixed angle $(w,\bw)$ on the asymptotic $S^2$. The first term is a linear ``soft'' photon (by which we mean momentum is strictly zero, as opposed to just small) contribution to the fixed-angle  charge. It does not contribute to the total charge $Q^+_1$ as it is a total derivative. The second term is the accumulated matter charge flux at the angle $(w,\bw)$. $Q^+_\ve$ generates the large gauge transformation on matter fields
\begin{equation}\label{qm}
\begin{split}
\left[   Q^+_\ve  , \Phi (u,z,\bz) \right] =
\left[   \int_{\ci^+} du' d^2 w \ve   \gamma_{w\bw} j_{u'}  , \Phi (u,z,\bz) \right] = - q \ve(z,\bz) \Phi (u,z,\bz) ,
\end{split}
\end{equation}
where $\Phi$ is any massless charged matter field operator on $\ci^+$ with charge $q$.

\section{Canonical formulation}\label{sec:canform}

The commutators, or equivalently a non-degenerate symplectic form, on the radiative phase space $\Gamma_R^+ \equiv \{ F_{uz}, F_{u\bz} \}$ are constructed, for example, in \cite{ash,frolov}. The non-vanishing ones are
\begin{equation}\label{cex}
\begin{split}
	\left[ F_{uz} (u, z, \bz), F_{u'\bw} ( u', w, \bw ) \right] =  \frac{ i e^2}{2} \p_{u} \delta \left( u - u' \right) \delta^2 ( z - w ) .
\end{split}
\end{equation}
Integrating and fixing the integration constants by antisymmetry gives
\begin{equation}
\begin{split}\label{dbracs}
	\left[ A_z (u, z, \bz) , A_\bw ( u', w, \bw ) \right] = - \frac{i e^2}{4} \Theta \left( u - u' \right) \delta^2 ( z - w ) ,
\end{split}
\end{equation}
where $\Theta(x) = \text{sign}(x)$. Given \eqref{qm}, one might expect that symmetry transformations on the gauge fields are then generated by commutators  with $Q^+_\ve$ using \eqref{dbracs}. However, an explicit computation gives
\begin{equation}
\begin{split}\label{gaugegenwrong}
	\left[ Q^+_\ve, A_z(u,z,\bz) \right] = \frac{i}{2} \p_z \ve (z,\bz) \neq i \delta_\ve A_z(u,z,\bz) ,
\end{split}
\end{equation}
which is off by a factor of $\frac{1}{2}$.  A similar factor of ${1\over 2}$ was encountered in the construction of the BMS supertranslation operator in \cite{hms}.

In order to resolve this discrepancy, we must give a more precise description of the phase space $\Gamma^+$. In particular we must both specify boundary conditions on $A_z$ at  the boundaries $\ci^+_\pm$ of $\ci^+$ and include the soft photon zero modes. The boundary values of the fields are denoted by
\begin{equation}
\begin{split}
A_z^\pm(z,\bz) \equiv A_z (u=\pm \infty, z,\bz)  .
\end{split}
\end{equation}
We consider here the sector of the phase space with no long-range magnetic fields, namely
\begin{equation}
\begin{split}\label{constrainteq}
	F_{z\bz} |_{\ci^+_\pm} = 0.
\end{split}
\end{equation}
In other words, the connections $A_z^\pm$ are flat on $\ci^+_\pm$. We will implement \eqref{constrainteq} as constraints. These constraints are not preserved by the commutators \eqref{dbracs}, which hence must be modified according to Dirac's procedure. A unique set of commutators are obtained by enforcing the continuity condition
\begin{equation}
\begin{split}\label{uniquebracketcond}
\left[  A^\pm_z (z,\bz) , A_\bw (u', w, \bw ) \right] &= \lim_{u\to\pm\infty} \left[ A_z (u, z, \bz )    ,  A_\bw (u', w, \bw ) \right]  , \\
\left[  A^+_z (z,\bz) -  A^-_z (z,\bz) , A^\pm_\bw (w, \bw ) \right] &= \lim_{u'\to\pm\infty}\left[  A^+_z (z,\bz) -  A^-_z (z,\bz) , A_\bw (u', w, \bw ) \right] , \\
\end{split}
\end{equation}
and the vanishing of equal -- $u$ commutators. The demand of continuity \eqref{uniquebracketcond} is not as innocuous as it looks because other commutators are \textit{not} continuous as $u$ is taken to the boundary. The choice \eqref{uniquebracketcond} gives an extension of the symplectic form on the radiative phase space to all of $\Gamma^+$ that is justified \text{a posteriori} by the fact that, as we now show, it leads to a realization of large gauge transformations as a canonical transformation on $\Gamma^+$. %We have not ruled out the possibility that there are inequivalent extensions of the symplectic form on the radiative phase space to all of $\Gamma^+$ corresponding to inequivalent quantizations of the boundary sector.

The solution to \eqref{constrainteq} is
\begin{equation}
\begin{split}\label{constsol1}
	A_z^\pm (z,\bz) = e^2 \p_z \phi_\pm(z,\bz) .
\end{split}
\end{equation}
Of course, the constant modes of $\phi_\pm$ cannot be determined from $A_z^\pm$, but it natural and useful to include them by simply treating $\phi_\pm (z,\bz)$ as unconstrained fields on $S^2$. The commutators satisfying \eqref{uniquebracketcond} are then
\begin{equation}
\begin{split}\label{dbrac1}
	\left[ \phi_{\pm}(z, \bz), A_w (u', w, \bw)\right]  &= \mp \frac{i}{8\pi}  \frac{1}{ z - w }  , \\
	\left[ \phi_+ (z, \bz), \phi_- (w, \bw)  \right]   &= \frac{i}{4\pi e^2 }\log \left| z - w \right|^2.  \\
\end{split}
\end{equation}
Using \eqref{constsol1}, the charge $Q^+_\ve$ can be written as
\begin{equation}
\begin{split}\label{charge1}
	Q^+_{\ve} &=  2 \int_{S^2} d^2 z \ve  \p_z \p_\bz \left( \phi_+ - \phi_- \right)  +  \int_{\ci^+} du d^2 z   \gamma_{z\bz} \ve j_u .
\end{split}
\end{equation}
It then immediately follows from \eqref{dbrac1} and \eqref{charge1} that
\begin{equation}
\begin{split}\label{gone}
	[ Q^+_\ve, A_z (u,z,\bz) ] &=  i \p_z \ve (z,\bz)  , \\
 	[ Q^+_\ve, \phi_\pm(z,\bz) ] &= \frac{i }{e^2} \ve(z,\bz) .
\end{split}
\end{equation}Moreover, the charges satisfy the Abelian algebra
\begin{equation}
\begin{split}
	\left[ Q^+_\ve, Q^+_{\ve'} \right] &= 0.
\end{split}
\end{equation}
Hence, on the constrained phase space defined by \eqref{constrainteq} the modified commutators properly generate the large gauge transformations.

Periodicity of $\ve$ implies that $\phi_-$ lives on a circle of radius $ 1 \over e^2$:
\be \phi_-\sim \phi_-+{ 2\pi \over e^2}.\ee
Exponentials of $\phi_-$ obey
\be \left[ Q^+_\ve,e^{ine^2\phi_-} \right] = -n \e e^{ine^2 \phi_-},\ee and have (in our conventions) integer charges $n$.  Such operators do not in themselves create physical states. Rather states with charge $n$ are created by products of these operators with neutral matter-sector operators. This is virtually the same operator product decomposition familiar in 2D CFT when factoring a $U(1)$ current algebra boson, or in 4D soft collinear effective field theory (SCET) involving the so-called jet field \cite{scet,mattilya}.

A vacuum wave function for the Goldstone mode which we take to be $\phi_-$ can be defined by the condition
\be\label{vc}\phi_-(z,\bz)|0\rangle =0.\ee
 \eqref{gone} implies that the large gauge symmetries are broken in this vacuum. The symmetries transform \eqref{vc} into   more general $\phi_-$ eigenstates obeying
\be \phi_-(z,\bz)|\alpha\rangle =\alpha(z,\bz)|\alpha\rangle.\ee
Up to an undetermined normalization, the inner products are
\be \langle \alpha |\alpha'\rangle =\prod_{z,\bz}\delta\left(\alpha(z,\bz)-\alpha'(z,\bz)\right). \ee
Other zero-energy states are
\be |\beta \rangle=\int\mathcal{D}\alpha \, e^{2i\int d^2z \p_z\alpha \p_\bz\beta} |\alpha\rangle.\ee
These are zero-mode eigenstates
\be \label{dsz} \int_{-\infty}^{+\infty} du F_{uz}|\beta \rangle =\p_z \beta |\beta \rangle \ee
obeying
\be  Q^+_\ve|\beta \rangle ={2\over e^2} \int d^2z\epsilon \p_z \p_\bz\beta |\beta \rangle.\ee
In particular, any state with $\beta=$ constant  has unbroken large gauge symmetry. These vacua are annihilated by the zero mode and are not the ones usually employed in QED analyses: it might be of interest to consider scattering in such states.\footnote{For example, such states might be related to the vacua considered in \cite{Kulish:1970ut,Ware:2013zja}. } Finally there are  normalizable, symmetry-breaking  vacua annihilated by complex linear combinations such as $\phi_-+i\phi_+$.

Let us restate and summarize this section. The classic expression \eqref{cex} is a non-degenerate symplectic form on the phase space of radiative modes with non-zero frequency. The soft photon, $i.e.$ zero frequency mode  $\int du F_{uz}=e^2\p_z(\phi_+-\phi_-)$,  is orthogonal to this form and has no symplectic partner among these radiative modes.  We remedy this by adding the boundary degree of freedom $A_z^-$ and constructing a symplectic form which pairs it with this zero mode. This is done consistently with the constraint \eqref{constrainteq} on both $A^-_z$ and the conjugate zero mode representing the absence of long range magnetic fields. The resulting non-degenerate phase space $\Gamma^+ = \left\{ F_{uz}(u,z,\bz),\phi_+(z,\bz),\phi_-(z,\bz)\right\}$ then consists of the usual (non-zero frequency) radiative modes, together with the zero-momentum soft photon $\phi_+(z,\bz)-\phi_-(z,\bz)$ and the  canonically conjugate periodic real boson $\phi_-(z,\bz)$.

\section{Asymptotic structure at $\ci^-$}\label{secim}

A similar structure exists near $\ci^-$ and is needed to discuss scattering. In this subsection we recap the requisite formulae. $\ci^-$ is at $r=\infty$ with $v$ fixed in advanced coordinates
\begin{equation}
\begin{split}\label{advancedcoord}
	ds^2 = - d v^2 + 2dv dr + 2r^2 \gamma_{z\bz} dz d\bz,
\end{split}
\end{equation}
which are  related to \eqref{retardedcoord} by the coordinate transfomation $u\to v-2r$, $r \to r$ and $z\to -1/\bz$.  The last relation means that points on $S^2$ with the same value of $z$ in retarded and advanced coordinates are antipodal. This coordinate choice is natural in the conformal compactification of Minkowski space, where there are null generators of $\ci$ which run from  $\ci^-$ to $\ci^+$ through spatial infinity.  We have chosen coordinates so that points anywhere on $\ci $ with the same value of $(z,\bz)$ lie on the same null generator.
In advanced radial gauge $\A_r=0=\A_v|_{\ci^-}$ and  the fields have the large $r$ expansion
\begin{equation}
\begin{split}\label{exp1}
	\A_z(r,v,z,\bz)&=B_z(v,z,\bz) + \Or(r^{-1}),~~~ \A_v (r, v,z,\bz) = \frac{1}{r} B_v (v, z,\bz) + \Or(r^{-2}) .
\end{split}
\end{equation}
The leading order constraint equation is (obtained from the $\mu = v$ component of the equations of motion)
\begin{equation}
\begin{split}\label{eom1}
	\gamma_{z\bz} \p_v B_v = -  \p_v \left( \p_z B_\bz + \p_\bz B_z \right) - e^2 \gamma_{z\bz}  j_v,
\end{split}
\end{equation}
where now
\begin{equation}
\begin{split}
	j_v (v,z,\bz) = \lim_{r\to\infty} \left[ r^2 j_v^M(v,r,z,\bz) \right] .
\end{split}
\end{equation}
Unfixed large gauge transformations are parameterized by $\ve^-(z,\bz)$ under which
\begin{equation}
\begin{split}
	\delta_{\ve^-} B_z=\p_z\e^-.
\end{split}
\end{equation}
The associated charge is
\begin{equation}
\begin{split}\label{charge2}
	Q^-_{\ve^-} &= -\frac{1}{e^2} \int_{\ci^-_+}d^2 z \gamma_{z\bz} \ve^- B_v = \frac{1}{e^2} \int_{-\infty}^\infty dv d^2 z \ve^- \left[ \p_v \left( \p_z B_\bz + \p_\bz B_z \right) + e^2 \gamma_{z\bz}  j_v \right] .
\end{split}
\end{equation}
As on $\ci^+$, we define the boundary values of the fields
\begin{equation}
\begin{split}
	B_z^\pm (z, \bz) \equiv B_z \left( v = \pm \infty, z, \bz \right)
\end{split}
\end{equation}
and impose constraints
\begin{equation}
\begin{split}\label{constrainteq1}
	\p_{[\bz}B^\pm_{z]} = 0.
\end{split}
\end{equation}
This is solved by
\begin{equation}
\begin{split}
	B_z^\pm = e^2 \p_z \psi_\pm .
\end{split}
\end{equation}
Employing the same methods and assumptions as our analysis near $\ci^+$, the commutators consistent with \eqref{constrainteq1} are
\begin{equation}
\begin{split}
	\left[ \psi_{\pm}(z, \bz), B_w (u', w, \bw)\right]  &= \mp \frac{i}{8\pi}  \frac{1}{ z - w }  ,  \\
	\left[ \psi_+ (z, \bz), \psi_- (w, \bw)  \right]   &= \frac{i}{4\pi e^2 }\log \left| z - w \right|^2   .  \\
\end{split}
\end{equation}
These in turn imply
\begin{equation}
\begin{split}\label{gtwo}
	[ Q^-_{\ve^-}, B_z (v,z,\bz) ] &=  i \p_z \ve^- (z,\bz)  , \\
 	[ Q^-_{\ve^-}, \psi_\pm(z,\bz) ] &= \frac{i }{e^2} \ve^-(z,\bz) .
\end{split}
\end{equation}
The incoming phase space is then $\Gamma^- = \left\{G_{vz}(v,z,\bz),\psi_+(z,\bz),\psi_-(z,\bz)\right\}$, where $G_{vz}=\p_vB_z$.

\section{Matching $\ci^+_-$ to $\ci^-_+$}\label{sec:matching}

The classical scattering problem is to find the map from $\Gamma^-$ to $\Gamma^+$, i.e. to determine the final data $\left(F_{uz},\phi_-\right)$ on \ip\ which arises from a given set of initial data $\left(G_{vz},\psi_+\right)$ on $\ci^-$.
Given a field strength everywhere on Minkowski space, this data is so far determined only up to the large gauge transformations which are generated by both $\ve$ and $\ve^-$ and act separately on $\Gamma^+$ and $\Gamma^-$.
Clearly there can be no sensible scattering problem without imposing a relation between $\ve$ and $\ve^-$.
Any relation between them should preserve Lorentz invariance. Under an $SL(2,{ \mathbb C } )$ Lorentz transformation parameterized by $\zeta^z \sim 1,z,z^2$ one finds
\begin{equation}
\begin{split}
\delta_\zeta \psi_+ = (\zeta^z\p_z+\zeta^\bz\p_\bz)\psi_+,  \\
\delta_\zeta \phi_- = (\zeta^z\p_z+\zeta^\bz\p_\bz)\phi_-.
\end{split}
\end{equation}
This symmetry is preserved by the natural requirement
\begin{equation}
\begin{split}\label{dsz}
\psi_+(z,\bz)=\phi_-(z,\bz)\,.
\end{split}
\end{equation}
\eqref{dsz} in turn requires
\begin{equation}
\begin{split}
\label{tty} \e(z,\bz)=\e^-(z,\bz),
\end{split}
\end{equation}
 as well as the generalization to finite gauge transformations.  Note that, because of the antipodal identification of the null generators of \ip\ and $\ci^-$,  this means the gauge parameter is $not$ the limit of a function which depends only on the angle in Minkowskian $(r,t)$ coordinates. Rather it goes to the same value at the beginning and end of light rays crossing through the origin of Minkowski space.\footnote{Such gauge transformations were considered in \cite{huot}.}
 $\e$ is then a function on the space of null generators of $\ci$.

Both the gauge field strength and the charge current are invariant under these symmetries.  The phases they generate on matter fields are classically unobservable. Hence (unlike the case of gravitational supertranslations considered in \cite{hms}), they have little import for the usual discussion of classical scattering.  It simply (antipodally) equates the final data for $\phi_-$ with that of the initial data for $\psi_+$. However in the quantum theory, where phases matter, they have significant consequences to which we now turn.

\section{Quantum Ward identity}\label{sec:quantum}

In this section, we consider the consequences of the large gauge symmetry on the semi-classical \cs-matrix. Let us denote an in (out) state comprised of $n$ ($m$) particles with charges $q^{\text{in}}_k$ ($q^{\text{out}}_k$), incoming at points $z_k^{\text{in}}$ (outgoing at points $z_k^{\text{out}}$) on the conformal sphere $S^2$ by $\ket{\text{in} } \equiv \ket{z_1^{\text{in} }, \cdots, z_n^{\text{in}}  }$ ($\bra{\text{out}} \equiv \bra{z_1^{\text{out} }, \cdots, z_m^{\text{out}}  }$). The \cs-matrix elements are then denoted as $\bra{\text{out}} {\cal S} \ket{ \text{in} }$.
The quantum version of the classical invariance of scattering under large gauge transformations is \begin{equation}\label{sm}
\begin{split}
\bra{\text{out}} \left( Q^+_\ve {\cal S} - {\cal S} Q^-_\ve  \right) \ket{ \text{in} }  = 0.
\end{split}
\end{equation}
The semi-classical charge obeys the quantum relations (from \eqref{charge} and \eqref{charge2})
\begin{equation}\label{xk}
\begin{split}
	\bra{z_1^{\text{out} },  \cdots, z_m^{\text{out}}  } Q^+_\ve &= \bra{z_1^{\text{out} },  \cdots, z_m^{\text{out}}  } F^+[\e]  +  \sum_{k=1}^m q^\text{out}_k \ve ( z_k^{\text{out}}, {\bar z}_k^{\text{out}} ) \bra{z_1^{\text{out} },  \cdots, z_m^{\text{out}}  }  , \\
	Q_\ve^- \ket{z_1^{\text{in} }, \cdots, z_n^{\text{in}}  } &= F^-[\e]\ket{z_1^{\text{in} }, \cdots, z_n^{\text{in}}  }  +   \ket{z_1^{\text{in} },   \cdots, z_n^{\text{in}}  } \sum_{k=1}^n q^\text{in}_k \ve ( z_k^{\text{in}}, {\bar z}_k^{\text{in}} )   ,
\end{split}
\end{equation}
where
\begin{equation}
\begin{split}
	F^+ [\ve] &\equiv  - 2 \int d^2 w \p_\bw \ve   \p_w\left( \phi_+ - \phi_-\right) , \\
	F^- [\ve]  &\equiv - 2 \int d^2 w  \p_\bw \ve \p_w \left( \psi_+ - \psi_- \right).
\end{split}
\end{equation}Defining
\begin{equation}
\begin{split}	\label{Fdef}
	F[\ve] \equiv F^+[\ve] - F^-[\ve]
\end{split}
\end{equation}
and  the time ordered product\begin{equation}
\begin{split}
 \colon F[\ve]  {\cal S}  \colon = F^+[\ve] {\cal S} - {\cal S} F^-[\ve] ,
\end{split}
\end{equation}  equation \eqref{sm} becomes  \begin{equation}
\begin{split}\label{weinsoftth}
	\bra{ \text{out} }  \colon F[\ve]  {\cal S}  \colon \ket{ \text{in} }  &= \left[\sum_{k=1}^n q^\text{in}_k \ve ( z_k^{\text{in}}, {\bar z}_k^{\text{in}}) - \sum_{k=1}^m q^\text{out}_k \ve ( z_k^{\text{out}}, {\bar z}_k^{\text{out}})  \right] \bra{ \text{out} }  {\cal S} \ket{ \text{in}} .
\end{split}
\end{equation}
This Ward identity relates the insertion of a soft photon with polarization and normalization given in \eqref{Fdef} into any  \cs-matrix element to the same \cs-matrix element without a soft photon insertion.

For an  incoming state which happens to be the vacuum, \eqref{xk} reduces to
\be \label{sag} Q_\ve^- \ket{0_{\text{in}}  } = F^-[\e]\ket{0_{\text{in}}  }.\ee Hence  $Q_\ve^-$ does not annihilate the vacuum unless
$\e={\rm constant}$, implying that all but the constant mode of the large gauge symmetries are spontaneously broken.
Moreover \eqref{gtwo} identifies $\psi_+$ as the corresponding Goldstone boson.

This result may seem surprising for the following reason. Soft photons are labelled by a spatial direction and a polarization. This suggests two modes for every point on the sphere, which is twice the number predicted by Goldstone's theorem. In fact the positive and negative helicity modes are not independent. As spelled out in Appendix \ref{appa}, there are non-local (on the asymptotic $S^2$) linear combinations of positive and negative helicity photons whose associated soft factor cancels exactly at leading order.\footnote{Subleading orders are considered in \cite{lps}.}  To leading order, these linear combinations of soft modes decouple from all \cs-matrix elements and hence are truly pure gauge. This relation reduces the two modes for every point on the sphere to the single one predicted by Goldstone theorem. Not accounting for this factor of two produced the wrong result in the charge commutator \eqref{gaugegenwrong} and was corrected for in the boundary condition in \eqref{constrainteq}.

\section{Soft photon theorem}\label{sec:softphoton}

 In this subsection, we show that the Ward identity \eqref{weinsoftth}  is the soft photon theorem in disguise.
In order to do so we must rewrite everything in momentum space.
The first step is to expand  the soft photon  operators $F^\pm[\ve]$, expressed above as weighted integrals over the conformal sphere at $\ci$,  in terms of the standard plane wave  in and out creation and annihilation operators.
Momentum eigenmodes in Minkowski space are usually described in flat coordinates
\begin{equation}
\begin{split}
	ds^2 = -dt^2 + d\vec{x} \cdot d \vec{x} ,
\end{split}
\end{equation}
related to the retarded coordinates in \eqref{retardedcoord} by
\begin{equation}
\begin{split}
	t = u + r, \quad x^1 + i x^2 = \frac{2rz}{1+z\bz}, \quad x^3 = \frac{r \left( 1 - z \bz \right)}{1 + z\bz },
\end{split}
\end{equation}
with $\vec{x} = (x^1, x^2, x^3)$ satisfying $\vec{x} \cdot \vec{x} = r^2$. At late times and large $r$, the wave packet for a massless particle with spatial momentum centered around $\vec{p}$ becomes localized on the conformal sphere near the point $(z,\bz)$ with
\begin{equation}\label{momentumrelabeling}
\begin{split}
	\vec{p} =  \omega {\hat x} = \frac{\omega}{1+z\bz} \left( z+ \bz , - i \left( z - \bz \right) , 1 - z \bz \right) ,
\end{split}
\end{equation}
where ${\hat x} = \frac{ \vec{x}}{r}$. The momentum of massless particles may be equivalently characterized either by $(\omega,z,\bz)$ or by $p^\mu$. At late times $t\to\infty$, the gauge field $\A_\mu$ becomes free and can be approximated by the mode expansion
\begin{equation}
\begin{split}\label{modexp}
	\A_\mu (x) = e \sum_{\alpha=\pm} \int \frac{d^3q}{(2\pi)^3} \frac{1}{2\omega_q} \left[ \ve_\mu^{\alpha^*} (\vec{q}) a_\alpha^{\text{out}} (\vec{q}) e^{i q \cdot x }  + \ve_\mu^{\alpha} (\vec{q}) a_\alpha^{\text{out}} (\vec{q})^\dagger e^{- i q \cdot x }  \right] ,
\end{split}
\end{equation}
where $q^0 = \omega_q = |\vec{q}|$ and $\alpha = \pm$ are the two helicities. The creation and annihilation operators on $\ci^+$, $a_\alpha^{\text{out}\dagger}$ and $a_\alpha^{\text{out}}$, obey
\begin{equation}
\begin{split}
	\left[ a_\alpha^{\text{out}} (\vec{q}) , a_\beta^{\text{out}} (\vec{q}')^\dagger \right] = \delta_{\alpha\beta} (2\pi)^3   (2 \omega_q) \delta^3 \left( \vec{q} - \vec{q}' \right)
\end{split}
\end{equation}
for $\omega_q > 0$ (for $\omega_q = 0$ the positive and negative helicities are linearly dependent; see Appendix \ref{appa}). Similarly, $a_\pm^{\text{in}}$ and $a_\pm^{\text{in}\dagger}$ annihilate and create incoming photons on $\ci^-$. In terms of $\omega, w$ and $\bw$, the momentum is
\begin{equation}
\begin{split}
q^\mu = \frac{\omega}{1+w\bw} \left( 1+ w\bw, w+\bw,-i(w-\bw),1-w\bw\right)
\end{split}
\end{equation}
and the polarization tensors have components
\begin{equation}
\begin{split}
	\ve^{+\mu}(\vec{q}) &= \frac{1}{\sqrt{2} } \left( \bw , 1 , - i , - \bw \right) , \\
	\ve^{-\mu}(\vec{q}) &= \frac{1}{\sqrt{2} } \left(w , 1 , i , -w \right), \\
\end{split}
\end{equation}
which satisfy $q_\mu \ve^{\pm \mu}(\vec{q}) = 0$.

To expand  the gauge field  near \ip\ recall that
\begin{equation}
\begin{split}
	A_z (u,z,\bz) = \lim_{r \to \infty} \A_z(u,r,z,\bz).
\end{split}
\end{equation}
Using $\A_z = \p_z x^\mu \A_\mu$, the mode expansion in \eqref{modexp} and the stationary phase approximation  we find
\begin{equation}
\begin{split}\label{modexp1}
	A_z (u,z,\bz) = - \frac{i}{8\pi^2}  \frac{\sqrt{2} e }{1+ z \bz }  \int_0^\infty d\omega_q \left[  a_+^{\text{out}} ( \omega_q {\hat x} ) e^{-i \omega_q u }   -  a_-^{\text{out}} ( \omega_q {\hat x} )^\dagger e^{i \omega_q u }   \right]  .
\end{split}
\end{equation}
Defining the energy eigenmodes
\begin{equation}
\begin{split}\label{nzdef}
	N_z^\omega (z,\bz) \equiv \int_{-\infty}^\infty du e^{i \omega u} F_{uz} ,
\end{split}
\end{equation}
we find
\begin{equation}
\begin{split}
	N_z^\omega (z,\bz) = - \frac{1}{4\pi}  \frac{\sqrt{2} e }{1+ z \bz }  \int_0^\infty d\omega_q \omega_q \left[  a_+^{\text{out}} ( \omega_q {\hat x} )  \delta \left( \omega - \omega_q \right)  + a_-^{\text{out}} ( \omega_q {\hat x} )^\dagger \delta \left( \omega + \omega_q \right)  \right]  .
\end{split}
\end{equation}
When $\omega > 0$ ($\omega < 0$) only the first (second) term contributes.  We define the zero mode by the hermitian expression
\begin{equation}
\begin{split}
	N_z^0 (z,\bz) = \lim_{\omega \to 0^+} \frac{1}{2} \left( N_z^\omega + N_z^{-\omega} \right) .
\end{split}
\end{equation}
It follows that
\begin{equation}
\begin{split}\label{ndef}
	N_z^0 (z,\bz) = -  \frac{1}{8\pi}  \frac{\sqrt{2} e }{1+ z \bz } \lim_{\omega \to 0^+ } \left[  \omega a_+^{\text{out}} ( \omega {\hat x} )  + \omega a_-^{\text{out}} ( \omega {\hat x} )^\dagger \right]  .
\end{split}
\end{equation}
Similarly on $\ci^-$ we define
\begin{equation}
\begin{split}\label{mdef}
	M_z^0 (z,\bz) \equiv \int_{-\infty}^\infty dv G_{vz} = - \frac{1}{8\pi}  \frac{\sqrt{2} e }{1+ z \bz } \lim_{\omega \to 0^+ } \left[  \omega a_+^{\text{in}} ( \omega {\hat x} )  + \omega a_-^{\text{in}} ( \omega {\hat x} )^\dagger \right]  ,
\end{split}
\end{equation}
where $a_\pm^{\text{in}}$ and $a_\pm^{\text{in}\dagger}$ annihilate and create incoming photons on $\ci^-$.

It follows from \eqref{nzdef} and \eqref{mdef} that
\begin{equation}
\begin{split}
	N_z^0 - M_z^0 = \int_{-\infty}^\infty du F_{uz}  - \int_{-\infty}^\infty dv G_{vz} = \frac{e^2}{4\pi} F \left[ \frac{1}{z-w} \right],
\end{split}
\end{equation}
where $F[\ve]$ is defined in \eqref{Fdef}. Setting $\ve(w,\bw) = \frac{1}{z-w}$, the Ward identity \eqref{weinsoftth} becomes
\begin{equation}
\begin{split}\label{weinsoftth1}
	\bra{ \text{out} }  \colon \left( N_z^0 - M_z^0 \right)   {\cal S}  \colon \ket{ \text{in} }  &= - \frac{e^2}{4\pi} \left[ \sum_{k=1}^m \frac{ q^\text{out}_k }{ z - z_k^{\text{out}} }     -  \sum_{k=1}^n \frac{ q^\text{in}_k }{ z - z_k^{\text{in}} }   \right] \bra{ \text{out} }  {\cal S} \ket{ \text{in} } .
\end{split}
\end{equation}
Using \eqref{ndef} and \eqref{mdef}, the above equations become
\begin{equation}
\begin{split}\label{weinbergsoftth}
 \lim_{\omega \to 0^+ } \left[ \omega \bra{ \text{out}} a_+^{\text{out}} ( \omega {\hat x} )    {\cal S} \ket{ \text{in} }  \right]  = \frac{e}{\sqrt{2}} (1+z\bz) \left[  \sum_{k=1}^m \frac{ q^\text{out}_k }{ z - z_k^{\text{out} } }  - \sum_{k=1}^n \frac{ q^\text{in}_k }{ z - z_k^{\text{in}} } \right] \bra{ \text{out} }  {\cal S} \ket{ \text{in} } ,
\end{split}
\end{equation}
where we have used the fact
\begin{equation}
\begin{split}
 \lim_{\omega \to 0^+ } \left[   \omega \bra{ \text{out} }   {\cal S}  a_-^{\text{in}} ( \omega {\hat x} )^\dagger \ket{ \text{in} }  \right]   = -   \lim_{\omega \to 0^+ } \left[ \omega \bra{ \text{out}} a_+^{\text{out}} ( \omega {\hat x} )    {\cal S} \ket{ \text{in} }  \right] .
\end{split}
\end{equation}

We now wish to compare \eqref{weinbergsoftth} with the soft photon theorem in its conventional form \cite{stevebook}
\begin{equation}
\begin{split}\label{spthstandform}
	\lim_{\omega \to 0^+ } \left[  \omega \cm_+ \left( p_\gamma ; \{p^{\text{in}}_k \}, \{p^{\text{out}}_k \}\right) \right] =
e \lim_{\omega \to 0^+}\left[  \sum_{k=1}^m \frac{ \omega q_k^{\text{out}} p^{\text{out}}_k  \cdot \ve^+ (p_\gamma ) }{ p^{\text{out}}_k  \cdot p_\gamma   }  - \sum_{k=1}^n \frac{ \omega q_k^{\text{in}}   p^{\text{in}}_k  \cdot \ve^+ (p_\gamma )  }{ p^{\text{in}}_k  \cdot p_\gamma   }  \right]  \cm \left(  \{p^{\text{in}}_k \}, \{p^{\text{out}}_k \}\right)  .
\end{split}
\end{equation}
Here, $\cm\left(  \{ p^{\text{in}}_k \}, \{p^{\text{out}}_k \}\right) $ is the momentum space scattering amplitude of $n$ ($m$) incoming (outgoing) particles with charges $q_k^{\text{in}}$ ($q_k^{\text{out}}$) and momenta $p^{\text{in}}_k $ ($p^{\text{out}}_k$) and $\cm_+ \left( p_\gamma ; \{p^{\text{in}}_k \}, \{p^{\text{out}}_k \}\right)$ is the same amplitude with one additional outgoing positive-helicity soft-photon with momentum $p_\gamma$. In our conventions
\begin{equation}
\begin{split}
 \cm \left(  \{p^{\text{in}}_k \}, \{p^{\text{out}}_k \}\right)  = \bra{ \text{out} }  {\cal S} \ket{ \text{in} },~~~ \cm_+ \left( p_\gamma ; \{p^{\text{in}}_k \}, \{p^{\text{out}}_k \}\right) = \bra{ \text{out}} a_+^{\text{out}} ( \vec{p}_\gamma)    {\cal S} \ket{ \text{in} }  .
\end{split}
\end{equation}
Using the parametrization of the momenta discussed earlier
\begin{equation}\label{wc}
\begin{split}
(p^{\text{in}}_k)^\mu &=  E^{\text{in}}_k \left(1 ,  \frac{ z^{\text{in}}_k + {\bar z}^{\text{in}}_k}{ 1 + z^{\text{in}}_k {\bar z}^{\text{in}}_k }, \frac{- i \left(  z^{\text{in}}_k - {\bar z}^{\text{in}}_k \right) }{ 1 + z^{\text{in}}_k {\bar z}^{\text{in}}_k }, \frac{1- z^{\text{in}}_k {\bar z}^{\text{in}}_k}{1+ z^{\text{in}}_k {\bar z}^{\text{in}}_k} \right)  , \\
( p^{\text{out}}_k)^\mu &=  E^{\text{out}}_k \left(1 ,  \frac{ z^{\text{out}}_k + {\bar z}^{\text{out}}_k}{ 1 + z^{\text{out}}_k {\bar z}^{\text{out}}_k }, \frac{- i \left(  z^{\text{out}}_k - {\bar z}^{\text{out}}_k \right) }{ 1 + z^{\text{out}}_k {\bar z}^{\text{out}}_k }, \frac{1- z^{\text{out}}_k {\bar z}^{\text{out}}_k}{1+ z^{\text{out}}_k {\bar z}^{\text{out}}_k} \right)  , \\
p_\gamma^\mu &=  \omega \left(1 ,  \frac{ z + {\bar z}}{ 1 + z {\bar z} }, \frac{- i \left(  z - {\bar z} \right) }{ 1 + z {\bar z} }, \frac{1- z {\bar z}}{1+ z {\bar z}} \right) , \\
{\bf \ve}^+_\mu(p_\gamma) &= \frac{1}{\sqrt{2}} \left( {\bar z}, 1, - i, - {\bar z} \right) ,
\end{split}
\end{equation}
we find
\begin{equation}
\begin{split}\label{theeq}
\lim_{\omega \to 0^+ }  \omega \cm_+ \left(p_\gamma; \{p^{\text{in}}_k \}, \{p^{\text{out}}_k \}\right) =
\frac{e}{ \sqrt{2} }  (1 + z \bz )  \left[  \sum_{k=1}^m   \frac{ q^{\text{out}}_k}{z - z_k^{\text{out}} } - \sum_{k=1}^n \frac{q^{\text{in}}_k}{z - z_k^{\text{in}} }  \right]   \cm \left(  \{p^{\text{in}}_k \}, \{p^{\text{out}}_k \}\right)  ,
\end{split}
\end{equation}
which is precisely \eqref{weinbergsoftth}. Thus, we have shown that the Ward identity, \eqref{weinsoftth} is equivalent to the soft-photon theorem. This argument can be run backwards to show that \eqref{spthstandform} implies \eqref{weinsoftth} with $\ve = \frac{1}{z-w}$. However, since any function $\ve(z,\bz)$ can be written as
\begin{equation}
\begin{split}
\ve(w,\bw) = \frac{1}{2\pi} \int d^2 z \ve(z,\bz) \p_\bz \frac{1}{z - w }
\end{split}
\end{equation}
and $F[\ve]$ is linear in $\ve$, the soft-photon theorem implies \eqref{weinsoftth} for any $\ve(z,\bz)$.

\section*{Acknowledgements}
We are grateful to F. Cachazo, T. Dumitrescu, I. Feige, D. Kapec, V. Lysov, J. Maldacena, S. Pasterski, A. Pathak, G. S. Ng, M. Schwartz and A. Zhiboedov for useful conversations. This work was supported in part by DOE grant DE-FG02-91ER40654 and the Fundamental Laws Initiative at Harvard.

\appendix

\section{Decoupled soft photons}\label{appa}

It is possible to see directly from the soft photon theorem that a particular combination of positive and negative helicity photons decouples from the theory. This is seen easiest in the $(z,\bz)$ coordinates. We start with the soft photon theorem in this parameterization \eqref{weinbergsoftth} for positive helicity insertions
\begin{equation}
\begin{split}\label{theeq1}
\lim_{\omega \to 0^+ }  \omega \bra{ \text{out}} a_+^{\text{out}} ( \omega {\hat x} )    {\cal S} \ket{ \text{in} }  =
\frac{e}{ \sqrt{2} }  (1 + z \bz )  \left[  \sum_{k=1}^m   \frac{ q^{\text{out}}_k}{z - z_k^{\text{out}} } - \sum_{k=1}^n \frac{q^{\text{in}}_k}{z - z_k^{\text{in}} }  \right]   \bra{ \text{out} }  {\cal S} \ket{ \text{in} } .
\end{split}
\end{equation}
Consider now the amplitude involving the following linear combination of the positive helicity soft photons
\begin{equation}
\begin{split}
{\cal O}(z,\bz) = \frac{1}{2\pi} (1+z\bz) \int d^2 w \frac{1}{\bz - \bw} \p_\bw \left[ { 1 \over 1+w \bw } \lim_{\omega \to 0^+} \left\{ \omega a_+^{\text{out}} (\omega {\hat y}) \right\} \right] ,
\end{split}
\end{equation}
where ${\hat y}$ points towards $(w,\bw)$. Insertions of this operator is given by \eqref{theeq1} as
\begin{equation}
\begin{split}
\bra{ \text{out}} {\cal O}(z,\bz)  {\cal S} \ket{ \text{in} }  =   \frac{e}{ \sqrt{2} }    (1+z\bz) \left[  \sum_{k=1}^m   \frac{ q^{\text{out}}_k}{\bz - \bz_k^{\text{out}}}   - \sum_{k=1}^n \frac{q^{\text{in}}_k}{\bz - \bz_k^{\text{in}} }    \right]   \bra{ \text{out} }  {\cal S} \ket{ \text{in} }  .
\end{split}
\end{equation}
This is precisely the soft photon theorem for a negative-helicity soft photon with momentum pointing towards $(z,\bz)$. We therefore conclude that the linear combination
\begin{equation}
\begin{split}
a_-^{\text{out}} (\omega {\hat x}) - \frac{1}{2\pi} (1+z\bz) \int d^2 w \frac{1}{\bz - \bw} \p_\bw \left[ {   a_+^{\text{out}} (\omega {\hat y})  \over 1+w \bw }  \right]
\end{split}
\end{equation}
has no poles and decouples from the \cs-matrix at leading order. In the more familiar momentum space variables, this is
\begin{equation}
\begin{split}
a_-^{\text{out}} (\omega {\hat p}_\gamma) + \frac{1}{2\pi \left( 1 + \cos\theta_{p_\gamma}  \right) } \int d\Omega_q \frac{1 + \cos\theta_q }{\left( \e^+({\hat p}_\gamma) \cdot {\hat q} \right)^2  }    a_+^{\text{out}} (\omega {\hat q} ) ,
\end{split}
\end{equation}
where the integral is over the angular distribution of ${\hat q}$.

\end{document}